\documentclass[aps,showpacs,superscriptaddress,twocolumn,amsmath,amssymb,prb]{revtex4}
\usepackage{graphicx}
\usepackage{dcolumn}
\usepackage{bm}
\usepackage{amssymb}
\usepackage{color}

\usepackage{hyperref}

\begin{document}

\title{Bistability and Hysteresis of Intersubband Absorption in Strongly Interacting Electrons on Liquid Helium}

\author{Denis Konstantinov}
\affiliation{Okinawa Institute of Science and Technology, Okinawa 904-0495, Japan}
\affiliation{ Low Temperature Physics Laboratory, RIKEN, Wako, Saitama 351-0198, Japan}
\author{M. I. Dykman}
\affiliation{ Department of Physics and Astronomy, Michigan State University, East Lansing, MI 48824, USA}
\author{	M.~J. Lea}
\affiliation{Department of Physics, Royal Holloway, University of London, TW20 0EX, United Kingdom}
\author{	Yu.~P. Monarkha}
\affiliation{ Institute for Low Temperature Physics and Engineering, Kharkov 61103, Ukraine}
\author{	K. Kono}
\affiliation{ Low Temperature Physics Laboratory, RIKEN, Wako, Saitama 351-0198, Japan}

\begin{abstract}
We study nonlinear inter-subband microwave absorption of electrons bound to the liquid helium surface. Already for a comparatively low radiation intensity, resonant absorption due to transitions between the two lowest subbands is accompanied by electron overheating. The overheating results in a significant population of higher subbands. The Coulomb interaction between electrons causes a shift of the resonant frequency, which depends on the population of the excited states and thus on the electron temperature $T_e$. The latter is determined experimentally from the electron photoconductivity. The experimentally established relationship between the frequency shift and $T_e$ is in reasonable agreement with the theory. The dependence of the shift on the radiation intensity introduces nonlinearity into the rate of the inter-subband absorption resulting in bistability and hysteresis of the resonant response. The hysteresis of the response explains the behavior in the regime of frequency modulation, which we observe for electrons on liquid $^3$He and which was previously seen for electrons on liquid $^4$He.
\end{abstract}
\date{\today}
\maketitle

\section{Introduction}

Free electrons outside liquid helium can be trapped at the vapor-liquid interface owing to the attractive polarization potential above the surface and a repulsive barrier at the surface, which prevents electrons from entering the liquid.~\cite{Cole1970,Andrei_book,Monarkha_book} As a result, the electron motion perpendicular to the helium surface is quantized, with energies $\epsilon_{\alpha}=-R/\alpha^2$ ($\alpha$=1,2,..), where $R$ is the effective Rydberg energy, $R \approx 7.6$~K for liquid $^4$He and $R \approx 4.2$~K for liquid $^3$He. An external electric field $E_{\perp}$ applied perpendicular to the surface Stark-shifts the energy levels. The shift depends on the quantum number $\alpha$. Consequently, the level spacing can be varied by changing $E_{\perp}$. If there are no other confining fields the electrons are free to move along the surface and their overall energy spectrum  consists of two-dimensional (2D) subbands.

Resonant inter-subband optical absorption due to transitions between the subbands has been directly observed for surface electrons (SEs) above liquid $^4$He using radiation in the millimeter \cite{Grimes1974,Collin2002} and far-infrared \cite{Lambert1980} ranges. Recently resonant absorption of millimeter-range microwaves was also observed for electrons on liquid $^3$He. \cite{Isshiki2007} In the experiments, SEs were tuned in resonance with radiation by adjusting the value of $E_{\perp}$ and optical absorption was detected as the variation in radiation power passing through the system.  Inter-subband absorption has been also probed indirectly by observing a change in the conductivity of irradiated SEs on both $^4$He and $^3$He. \cite{Volodin1981,KonstantinovJLTP2007} This change is mainly caused by radiation-induced overheating of the electron system.\cite{Konstantinov2007} The overheating occurs both where electrons are scattered primarily by helium vapor atoms or by surface vibrational modes, ripplons.~\cite{Monarkha2007,KonstantinovJPSJ2008,KonstantinovJLTP2008,KonstantinovLTP2008}

The recent interest in resonant absorption was stimulated by the proposal to use laterally confined electrons on the helium surface as qubits for quantum computing, with the qubit states being the out-of-plane states with $\alpha=1$ and 2.~\cite{Platzman1999,Dykman2003} Resonant interaction with an externally applied microwave field provides a natural way to manipulate the qubit state. Coupling between the qubits comes from the Coulomb interaction between the electrons. The interaction energy depends on the electron quantum numbers $\alpha$, because the electron distance from the surface depends on $\alpha$. Therefore, the change of the quantum state of one electron results in the shifts of the quantum levels of other electrons.

In microwave absorption experiments with electrons that are free to move along the surface, one would expect to see a shift in the resonance frequency for the transition from the ground to the first excited subband (and also in other transition frequencies) depending on the population of excited subbands. The subband-population dependent shifts of the energy levels due to many-electron effects were predicted earlier for semiconductor heterostructures. \cite{Allen1976,Ando1977} In such systems the electron correlation effects are a perturbation, because the system is close to metallic. 

A distinctive feature of the electron system on helium is that it is strongly correlated.\cite{Andrei_book,Monarkha_book,Dykman1979,Dykman1993,Monarkha1997,Lea1998} Therefore the frequency shifts turn out to be much stronger than in semiconductors. Signs of the dependence of the frequency shift on the state of the system were seen earlier in a complicated response to radiation with modulated frequency\cite{Glasson2004} and the power-dependent shift of the photoconductivity signal.\cite{KonstantinovJPSJ2008} However, no systematic studies of these effects have been carried out.

The state-dependent shift of resonance frequency introduces nonlinearity in the absorption rate. Such nonlinearity can cause bistability and hysteresis of resonant response. Hysteretic effects related to nonlinear absorption are familiar already from the early work on ferromagnetic resonance in ferrite crystal disks.\cite{Bloembergen1954,Anderson1955,Weiss1958} For 2D systems, magnetization bistability was observed in spin-polarized atomic hydrogen gas absorbed on the superfluid helium film.\cite{Vasilyev2002}.  For semiconductor heterostructures, the nonlinearity of inter-subband absorption was predicted and studied since late 80s \cite{Newson1987,Zaluzny1993,Craig1996,Lutgen1996,Shtrichman2001,Muller2004} and absorption bistability has been long sought.\cite{Stockman1993,Batista2002,Wijewardane2004,Li2007}

In this work, we present theoretical and experimental studies of nonlinear absorption effects for a strongly correlated electron system on liquid helium surface.  A short account of the results appeared earlier. \cite{KonstantinovPRL2009} We provide a theory of the frequency shift associated with electron correlations and show that this shift causes bistability and hysteresis of resonant response. We describe experimental results on the frequency shift and nonlinear absorption effects for electrons on liquid $^3$He in the temperature range 0.2$-$0.4~K that covers both vapor-atom and ripplon scattering regimes. The response is observed as a change of the longitudinal conductivity $\sigma_{xx}$ of SEs in a weak perpendicular magnetic fields. The frequency shift is measured as a function of electron temperature $T_e$, which is determined from the relative change of $\sigma_{xx}$ at resonance. The shift is compared with the theoretical estimate and a reasonable agreement is found.

At high values of input microwave power, the absorption cross-section shows a jump as a function of $E_{\perp}$ and depends on the direction of sweeping $E_{\perp}$, that is, displays hysteresis. This behavior agrees well with our theoretically predicted bistability of nonlinear response. We also describe the results of the experiment, in which resonance is observed by direct measurements of the microwave power absorbed by the electron system. We show that the complicated behavior of the response to radiation with slowly modulated frequency can be explained by taking into account the nonlinearity of the absorption rate.

In Sec.~II we derive a many-body quantum kinetic equation for a strongly correlated system and use it to describe nonlinear resonant absorption and to predict bistability of resonant response. In Sec.~III we describe the experimental techniques. In Sec.~IV the experimental results are presented. In Section~V we discuss the results of the measurements and make comparison with the theory. The experimentally observed nonlinear effects, in particular the bistability and hysteresis of inter-subband absorption, are also discussed in this section. Section~VI provides a summary of the main results and the conclusions.
\section{Theoretical background}

\subsection{Qualitative picture of electron dynamics}

In the neglect of the electron-electron interaction, electron motion normal to the helium surface is described by the Schr\"odinger equation
\begin{equation}
-\frac{\hbar^2}{2m}\frac{d^2}{dz^2}\psi_{\alpha}(z)+V(z)\psi_{\alpha}(z)=\epsilon_{\alpha}\psi_{\alpha}({z}),
\label{eq:1}
\end{equation}
where $z$ is the out-of-plane coordinate, $\psi_{\alpha}(z)$ and $\epsilon_{\alpha}$ are the eigenfunctions and eigenvalues ($\alpha=1,2,\ldots$), $m$ is the electron mass, and $V(z)$ is the electron potential energy, $V(z)=-\Lambda/z+eE_{\perp}z$ for $z>0$. Here, $\Lambda=e^2(\varepsilon-1)/4(\varepsilon+1)$, $-e$ is the electron charge, $\varepsilon$ is the dielectric constant of liquid helium, and $E_{\perp}$ is the electric field applied normal to the surface. On the surface, $z=0$, the potential $V(z)$ has a high step $\sim 1$~eV. Neglecting penetration of the electron wave function into the liquid, Eq.~(\ref{eq:1}) can be solved numerically with the boundary condition $\psi_{\alpha}(0)=0$. For the field $E_{\perp}<10^2$~V/cm, the typical localization length of the low-lying electron states is $r_{\rm B}=\hbar^2/\Lambda m\approx 0.8\times 10^{-6}$~cm$^{-1}$ for $^4$He.

Of central interest for this paper is electron response to a microwave field which is resonant with the $\alpha=1 \to \alpha=2$ transition. This response depends on electron relaxation. To describe the relaxation we note that, for electron densities $n_s$ and temperatures $T$ studied in experiment, the electron-electron interaction is strong. The interaction is characterized by the plasma parameter $\Gamma$ and the characteristic plasma frequency $\omega_p$,
\begin{equation}
\Gamma=e^2(\pi n_s)^{1/2}/k_{\rm B} T, \quad \omega_p=(2\pi e^2 n_s^{3/2}/m)^{1/2}.
\label{eq:2}
\end{equation}
Most of the existing experimental results on electrons on helium refer to the range $\Gamma \gg 1$. This means that the electron system is strongly correlated, SEs form a nondegenerate liquid or, for $\Gamma >130$, a Wigner solid; the experiment described in this paper refers to the range where the system is a liquid. The reciprocal plasma frequency $\omega_p^{-1}$ gives the characteristic time of inter-electron in-plane momentum and energy exchange. In our experiment $\omega_p>6\times 10^9$~s$^{-1}$. This was the shortest relaxation time in the range of temperatures studied in our experiment. Also, $\omega_p$ exceeded the microwave absorption rate, see below.

The fast inter-electron energy exchange leads to a thermal distribution over the in-plane electron energy, with an effective electron temperature $T_e$. We note that the time it takes to form such distribution in a correlated electron system can be longer than $\omega_p^{-1}$. For example, if one thinks of electrons forming a Wigner crystal, this is the thermalization time of the phonons of the Wigner crystal. Still this time is much shorter than the characteristic energy relaxation time due to inelastic electron scattering by helium vapor atoms or ripplons/phonons in helium. Of major interest to us is the parameter range where $k_{\rm B} T_e>\hbar\omega_p$. In this range electron in-plane motion is semi-classical and the electron in-plane state can be described by well-defined momentum and coordinate.

The temperature $T_e$ should be the same in all subbands, even though the distribution over the subbands does not necessarily has to be thermal. This is a consequence of the strong difference between the characteristic in-plane and out-of-plane inter-electron distance in the correlated system, $n_s^{-1/2}$ and $r_B$. Since $n_s^{-1/2}\gg r_B$, the in-plane momentum exchange $\Delta p_{ee}$ is essentially the same whether the electrons are in the same or in different subbands, leading to the same in-plane momentum distribution irrespective of the subband. At the same time, the electron-electron interaction practically does not lead to inter-subband transitions, because it requires short-range collisions which do not occur in the correlated system.

Other scattering mechanisms come from the interaction of SEs with the surrounding, primarily with helium vapor atoms, for comparatively high temperatures, $T>0.3$~K for electrons on $^3$He, and with surface waves and bulk excitations in liquid helium, for lower temperatures.\cite{Monarkha_book,Platzman1999} Scattering by vapor atoms and by surface capillary waves, ripplons, is mostly quasi-elastic. This is a consequence, respectively, of the large ratio of the helium atom mass to the electron mass and the slowness of ripplons: even for the ripplon wave number as large as $r_{\rm B}^{-1}$ the ripplon frequency is very small, $\sim 10^8$~Hz, and the ripplon energy is much smaller than the average energy of in-plane electron motion.

\subsection{Hamiltonian of the many-electron system}

Using the condition $n_s^{-1/2}\gg r_B$, the Hamiltonian of the isolated electron system can be written as $H_0+H_{ee}^{(1)}$, where the leading term is
\begin{eqnarray}
&& H_0=H_K + H_{ee}^{(0)}+H_z, \quad H_K=\frac{1}{2m}\sum\limits_n {\bf p}_n^2, \nonumber\\
&& H_{ee}^{(0)}=\frac{e^2}{2}\sum\limits_{n\neq n'} r_{nn'}^{-1}, \qquad H_z=\sum\limits_{n,\alpha} \epsilon_{\alpha} \sigma_n^{\alpha\alpha}. 
\label{eq:3}
\end{eqnarray}
Here, ${\bf r}_n$ and ${\bf p}_n$ ate lateral coordinate and momentum of an $n$th electron, respectively, $r_{nn'}\equiv |{\bf r}_n - {\bf r}_{n'}|$ is the in-plane inter-electron distance, and
\begin{equation}
\sigma_n^{\alpha\beta}=|\alpha\rangle_n\,_n\!\langle \beta|,
\label{eq:33}
\end{equation}
where $|\alpha\rangle_n \equiv [\psi_{\alpha}]_n$ is the $\alpha$th out-of-plane state of the $n$th electron.

The part of the electron-electron interaction incorporated into $H_0$ corresponds to the approximation of the electrons being in a plane. The part of the interaction related to the out-of-plane electron displacement is described by $H_{ee}^{(1)}$. To the leading order in $r_{\rm B}/r_{nn'}$
\begin{equation}
H_{ee}^{(1)}=-\frac{e^2}{4}\sum\limits_{n\neq n'}(z_n - z_{n'})^2/r_{nn'}^3.
\label{eq:4}
\end{equation}
As we will see, this interaction leads to an analog of the depolarization effect in semiconductor heterostructires.\cite{Allen1976,Ando1977,Zaluzny1993,Craig1996,Lutgen1996,Shtrichman2001,Muller2004} However, because SEs on helium are strongly correlated the effect is much stronger and is described in a qualitatively different way.

The electron coupling to a resonant microwave field is described by the Hamiltonian
\begin{equation}
H_F=-\hbar\Omega_R {\rm cos}\omega_F t\sum_n \big( \sigma_n^{12} + \sigma_n^{21} \big).
\label{eq:5}
\end{equation}
Here, $\Omega_R=eE_{\rm MW}z_{12}/\hbar$ is the Rabi frequency; $E_{\rm MW}$ and $\omega_F$ are the amplitude and frequency of the radiation field, respectively, and $z_{\alpha\beta}=\langle\alpha|z|\beta\rangle$. Frequency $\omega_F$ is assumed to be close to the transition frequency of the electron system, $\omega_{21}=(\epsilon_2-\epsilon_1)/\hbar$, $|\omega_{21}-\omega_F|\ll \omega_{21}$.

The nonlinear response of a strongly correlated system can be analyzed using the many-electron density matrix. It depends on the states of in-plane and out-of-plane motion of all electrons. It is convenient to write it in the interaction representation using canonical transformation $U(t)={\rm exp}(-iH_tt)$, where
\begin{equation}
H_t=H_0+\hbar(\omega_F-\omega_{21})\sum_n\sigma_n^{22} + H_{He}.
\label{eq:6}
\end{equation}
The term $H_{He}$ in Eq.~(\ref{eq:6}) describes excitations in liquid helium and the almost ideal gas formed by helium vapor atoms. The transformed Hamiltonian of coupling to the field $U^{\dagger}H_F U$ is independent of time in the rotating wave approximation. We emphasize that the electron-electron interaction $H_{ee}^{(0)}$ is incorporated into $H_t$, this is not a perturbation.

We write the density matrix in the interaction representation with Hamiltonian $H_t$ using the Wigner representation with respect to in-plane motion as $\rho_0(t;\{\alpha_n,\beta_n,{\rm r}_n,{\rm p}_n\})$. For weak coupling to ripplons and helium vapor atoms, one can obtain a many-electron kinetic equation for $\rho_0$. This equation is Markovian in time slow compared to $\hbar/k_BT_e, \omega_{21}^{-1}$ and presents an immediate extension to the multi-subband case of the kinetic equation discussed previously in the one-subband approximation.\cite{Dykman1997} 

Because the Coulomb interaction is strong, to zeroth order in the interaction with excitations in helium and in $H_{ee}^{(1)}$ function $\rho_0$ depends on the coordinates and momenta of individual electrons in terms of the total in-plane momentum ${\bf P}=\sum_n {\bf p}_n$ and the total in-plane energy $H_K+H_{ee}^{(0)}$. The distribution over the total in-plane energy is of the Boltzmann form. Moreover, in the classical range $k_{\rm B}T_e>\hbar\omega_p$ the distribution over momenta ${\bf p}_n$ is also of the Maxwell form.

Coupling to helium vapor atoms and ripplons causes mixing of states with different ${\bf P}$ and also of different out-of-plane states $|\alpha\rangle_n$. The coupling Hamiltonian can be written as
\begin{equation}
H_i=\sum\limits_{{\bf q},n}\sum\limits_{\alpha,\beta}\hat{V}_{{\bf q}\alpha\beta}e^{i{\bf qr}_n}\sigma_n^{\alpha\beta}.
\label{fig:7}
\end{equation}
Here, $\hat{V}$ is an operator with respect to the variables of ripplons or phonons in helium and the positions of the helium vapor atoms, while $\hbar{\bf q}$ is the lateral momentum transferred from these excitations to an electron. Terms with $\alpha=\beta$ describe intrasubband coupling, whereas terms with $\alpha\neq \beta$ describe mixing of states in different subbands.

The characteristic values of the transferred momentum are comparatively large, $q\gg n_s^{-1/2}$, because the density of states of excitations in helium and often the interaction strength increase with $q$. Thus scattering by helium excitations is short-range. As a result, in the electron liquid each electron is scattered individually, processes where two electrons are scattered by the same ripplon, phonon, or a vapor atom can be disregarded.\cite{Dykman1979,Dykman1997} Since between the scattering events the electrons have time to exchange lateral momentum and energy with each other, the scattering rates can be averaged over lateral electron motion with the Boltzmann factor.

Resonant response of the electron system to microwave field is independent of the electron lateral motion. Of interest for the study of this response is the single-particle density matrix
\begin{equation}
\rho_{\alpha\beta}(t)={\rm Tr}\left[ \sigma_n^{\beta\alpha}\hat\rho_0 (t)\right]=N^{-1}\sum_n{\rm Tr}\left[ \sigma_n^{\beta\alpha}\hat\rho_0 (t)\right].
\label{eq:8}
\end{equation}
The trace is taken over the in-plane coordinates and momenta and over the out-of-plane states of all electrons, $N$ is the total number of electrons. In a spatially uniform system $\rho_{\alpha\beta}$ is independent of the subscript $n$ in Eq.~(\ref{eq:8}).

\subsection{Many-electron shift of the transition frequency}

Qualitatively, the major effect of the interaction $H_{ee}^{(1)}$ on resonant absorption is the linear Stark shift of the frequency of the inter-subband $|1\rangle\rightarrow |2\rangle$ transition of an electron due to the out-of-plane component of the electric field created by other electrons. This field for an $n$th electron is given by $e^{-1}\partial H_{ee}^{(1)}/\partial z_n$. As seen from Eq.~(\ref{eq:4}), the field depends on the relative distance between the electrons normal to the surface and therefore on the states of out-of-plane motion of different electrons, as well as on the inter-electron distance in the plane. We emphasize that we are interested in the field on an electron, not in the average field in the electron layer, which is significantly different.

Formally, the term $H_{ee}^{(1)}$ couples a single-particle density matrix, in particular that given by Eq.~(\ref{eq:8}), to a two-particle one. This coupling is described by the term in the kinetic equation $(\partial_t\hat{\rho}_0)_{ee} = i\hbar^{-1}\left[ \hat{\rho}_0(t),H_{ee}^{(1)}(t)\right]$, where $H_{ee}^{(1)}(t) = U^{\dagger}(t)H_{ee}^{(1)}U(t)$. The density matrix in the interaction representation slowly varies in time. Therefore the main contribution to $\hat{\rho}_0$ comes from slowly varying terms in $H_{ee}^{(1)}(t)$. 

To find the smooth terms $H_{ee}^{(1)}(t)$ we write in Eq.~(\ref{eq:4}) $z_n=\sum_{\alpha\beta} z_{\alpha\beta}\sigma_n^{\alpha\beta}$ (and similarly for $z_n^2$) and notice that $U^{\dagger}(t)\sigma_n^{\alpha\beta}U(t)\propto {\rm exp}(i\omega_{\alpha\beta}t)$. Here $z_{\alpha\beta}=_n\!\!\langle\alpha|z_n|\beta\rangle_n$; the matrix elements of $z_n^2$ are defined similarly. The inter-subband transition frequencies $\omega_{\alpha\beta}$ largely exceed the in-plane vibration frequencies, which are $\sim\omega_p$ and characterize time variation of the factors $r_{nn'}^{-3}$  in $H_{ee}^{(1)}(t)$. Therefore the first type of slowly varying matrix elements in $H_{ee}^{(1)}(t)$ are those containing only diagonal components $z_{\alpha\alpha}\sigma_n^{\alpha\alpha}$ and $(z^2)_{\alpha\alpha}\sigma_n^{\alpha\alpha}$. The second type comes from the terms $\propto z_nz_{n'}$ with $n\neq n'$. The corresponding slowly varying terms are proportional to $|z_{\alpha\beta}|^2\sigma_n^{\alpha\beta}\sigma_{n'}^{\beta\alpha}$ with $\alpha\neq \beta$.

We will analyze the interaction $H_{ee}^{(1)}$ in the mean-field approximation. This approximation is justified when the number of nearest neighbors for each electron is large, so that correlations between out-of-plane states of neighboring electrons and their positions are averaged out. In other words, tracing over the positions of coupled electrons and over their out-of-plane states is done independently. In a strongly correlated 2D electron liquid the number of nearest neighbors is 6, on average, which makes the approximation reasonable. For typical times on the order of the duration of a collision with helium excitations or a radiation-induced inter-subband transition the relative shift of the positions of neighboring electrons is small. It is a good approximation then to describe the distribution of inter-electron distances $r_{nn'}$ by the static pair correlation function $g(r_{nn'})$.

The many-electron frequency shift $\Delta \omega_{\beta\alpha}$ of the $|\alpha\rangle \to |\beta\rangle$ transition is determined by the ratio $\hbar^{-1}{\rm Tr}\left\{ \sigma_n^{\alpha\beta}\big[H_{ee}^{(1)},\hat{\rho}_0\big] \right\}/\rho_{\beta\alpha}$. From the above arguments
\begin{eqnarray}
{\rm Tr} \left\{ \sigma_n^{\alpha\beta}\sum\nolimits_{n'} 'r_{nn'}^{-3}\hat{\rho}^{(0)} \right\}\approx Fn_s^{3/2}\rho_{\beta\alpha}, \nonumber\\
{\rm Tr} \left\{ \sigma_n^{\alpha\beta}\sum\nolimits_{n'} ' r_{nn'}^{-3}\sigma_{n'}^{\alpha '\beta '}\hat{\rho}^{(0)} \right\}\approx Fn_s^{3/2}\rho_{\beta\alpha}\rho_{\beta'\alpha'},
\label{eq:9}
\end{eqnarray}
where the prime over the sum indicates that $n'\neq n$ and
\begin{equation}
F=n_s^{-3/2} \Big \langle \sum\nolimits_{n'} 'r_{nn'}^{-3} \Big \rangle.
\label{eq:10}
\end{equation}
Numerically, in a strongly correlated liquid $F\approx 8.91$~\cite{Fang1997}.

From Eqs.~(\ref{eq:4}) and (\ref{eq:9}) with account taken of the commutation relation $\left[ \sigma_n^{\alpha\beta},\sigma_{n'}^{\alpha'\beta'} \right]=\delta_{nn'}\left( \sigma_n^{\alpha\beta'}\delta_{\beta',\alpha'} - \sigma_n^{\alpha'\beta}\delta_{\alpha,\beta'} \right)$, the frequency shift of the $|1\rangle \to |2\rangle$ transition due to the electron-electron interaction is
\begin{eqnarray}
&&\Delta\omega_{21}=\frac{Fe^2n_s^{3/2}}{2\hbar} \Big[ (z^2)_{11}-(z^2)_{22} - 2\big( z_{11}- z_{22}\big) \nonumber \\
&& \times \sum\nolimits_\alpha z_{\alpha\alpha}\rho_{\alpha\alpha} + 2|z_{12}|^2 \big( \rho_{11}-\rho_{22}\big)\Big].
\label{eq:11}
\end{eqnarray}
From Eq.~(\ref{eq:11}), the frequency shift $\Delta\omega_{21}$ is determined by the populations of all subbands $\rho_{\alpha\alpha}$. In Eq.~(\ref{eq:11}) we dropped terms proportional to $\rho_{1\alpha}\rho_{\alpha 2}$ with $\alpha\neq 1,2$. These terms would be important if radiation excited transitions other than $|1\rangle \to |2\rangle$. However, such transitions are nonresonant, and for a comparatively small microwave power that we consider they can be disregarded.

The effect of the electron correlations on the position of the absorption line was considered earlier by Lambert and Richards \cite{Lambert1980}. Their results referred to the weak-power limit, $\rho_{\alpha\alpha}\propto \delta_{\alpha,1}$, and the expression they were using differed from Eq.~(\ref{eq:11}); in particular, the last term was missing all together.  

The physics of the terms proportional to the diagonal and off-diagonal matrix elements of $z$, $z^2$ in Eq.~(\ref{eq:11}) is different. The diagonal terms come from the fact that SEs in different out-of-plane states $|\alpha\rangle$ have different static dipole moments in the out-of-plane direction. These static moments lead to the static Stark shift of the transition frequency $\omega_{21}$ of other electrons, which depends on the occupancies $\rho_{\alpha\alpha}$. The term proportional to $|z_{12}|^2$ in Eq.~(\ref{eq:11}) is dynamic in nature: it describes resonant excitation transfer between different electrons.\cite{Platzman1999}

\subsection{Kinetic equation}

An equation for the single-electron density matrix for out-of-plane motion $\rho_{\alpha\beta}$ can be obtained from the full kinetic equation for $\hat\rho_0$ in the interaction representation,\cite{Ryvkin}
\begin{eqnarray}
\label{eq:12}
&& \dot{\rho}_{\alpha\alpha}=\sum_{\beta}\left(\gamma_{\beta\alpha}\rho_{\beta\beta}- \gamma_{\alpha\beta}\rho_{\alpha\alpha}\right) +\Omega_R \left(\delta_{\alpha,1}-\delta_{\alpha,2}\right) {\rm Im} \rho_{12}, \nonumber \\
&& \dot{\rho}_{12}=-(i\delta\omega + \gamma_0)\rho_{12} + \frac{1}{2}i\Omega_R(\rho_{22}-\rho_{11}),
\end{eqnarray}
where
\begin{equation}
\label{eq:Delta_omega}
\delta\omega = \omega_F-\omega_{21}-\Delta\omega_{21}.
\end{equation}
In Eq.~(\ref{eq:12}) we do not incorporate explicitly the electron-electron interaction, which leads to establishing thermal distribution over lateral motion, with effective temperature $T_e$. We also do not incorporate the relaxation processes that lead to energy exchange between the electrons and the helium excitations. 

Parameters $\gamma_{\alpha\beta}$ in Eq.~(\ref{eq:12}) are the rates of inter-subband transitions $|\alpha\rangle \to |\beta\rangle$, which are essentially quasi-elastic. The electron momentum goes to helium excitations whereas the energy of motion normal to the helium surface goes into the kinetic energy of lateral motion and the energy of the electron-electron interaction,
\begin{eqnarray}
\label{eq:13}
\gamma_{\alpha\beta}
&=&\hbar^{-2}\sum_{\bf q}\overline{ |V_{{\bf q}\alpha\beta}|^2}\int_{-\infty}^{\infty}dt \exp[i(\epsilon_{\alpha}-\epsilon_{\beta} )t/\hbar]\nonumber\\
&&\times\left\langle \exp[i{\bf qr}_n(t)]\exp[-i{\bf qr}_n(0)]\right\rangle.
\end{eqnarray}
Here, the overline indicates averaging over the thermal distribution (with temperature $T$) of the relevant excitations in helium; $\langle\ldots\rangle$ denotes averaging over the in-plane electron motion with temperature $T_e$. Expression (\ref{eq:13}) is independent of the electron number $n$. In deriving it we used the aforementioned picture in which electrons are scattered by a short-range potential independently, that is one electron collides with a short-range scatterer at a time. However, during a collision the electron is in the in-plane potential created by other electrons, which is determined by $H_{ee}^{(0)}$. This potential is smooth on the electron wavelength $\lambda_T$ and for $k_BT_e\gg \hbar\omega_p$ weakly affects short-range scattering in the absence of a magnetic field.\cite{Dykman1997}

The electron-electron interaction plays a critical role if the electron system is in a strong magnetic field $B$ normal to the surface. Here, in the single-electron approximation the electron energy spectrum is discrete (the Landau quantization), and the Born approximation used to derive Eq.~(\ref{eq:13}) would not apply. However, because an electron is in an a fluctuational electric field of other electrons $E_{\rm f}$, its potential (and thus kinetic) energy is uncertain by $eE_{\rm f}\lambda_T\sim \hbar\omega_p$, and if this uncertainty exceeds $\hbar\omega_c$, where   $\omega_c=eB/mc$ is the cyclotron frequency, the effect of the Landau quantization on scattering is essentially eliminated.\cite{Dykman1993} This condition was met in the experiment, even though the magnetic field was classically strong. 
\begin{figure}[t]
\centering
\includegraphics[width=6.5cm]{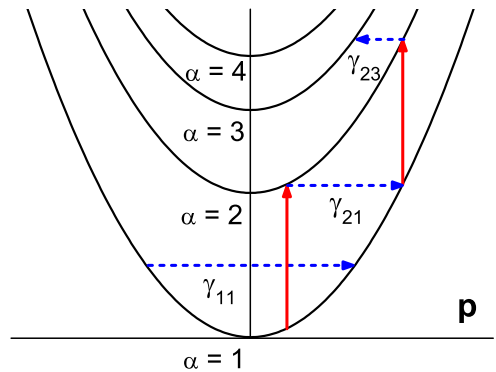}
\caption{(color online) Sketch of the energy subbands and the electron transitions. Vertical (red) arrows indicate photon-induced transitions, while horizontal (blue) arrows indicate transitions due to elastic scattering by excitations in helium.}
\label{fig:1}
\end{figure}

The processes described by Eq.~(\ref{eq:13}) are shown schematically in Fig.~\ref{fig:1}. The calculation of the rates $\gamma_{\alpha\beta}$ is similar to that of the in-plane relaxation rate in a correlated electron system.\cite{Dykman1997} Even where $T_e\neq T$, since the scattering described by Eq.~(\ref{eq:13}) is elastic and the electron distribution is thermal, the transition rates satisfy the detailed balance condition
\begin{equation}
\gamma_{\alpha\beta}=\gamma_{\beta\alpha} \exp [(\epsilon_{\alpha}-\epsilon_{\beta})/k_{\rm B} T_e].
\label{eq:14}
\end{equation}

The decay rate $\gamma_0$  of the off-diagonal matrix element $\rho_{12}$ in Eq.~(\ref{eq:12}) in the single-electron approximation for $T_e=T\ll (\epsilon_2-\epsilon_1)\hbar$ was considered in Ref.~\onlinecite{Ando1978}. Using the approach of Ref.~\onlinecite{Dykman1997} the results can be extended to the many-electron case and made applicable in the presence of a magnetic field normal to the helium surface, 
\begin{eqnarray}
\label{eq:15}
 \gamma_0=&&\frac{1}{2} (\sum_{\alpha\neq 1}\gamma_{1\alpha}+\sum_{\alpha\neq 2}\gamma_{2\alpha}) + \frac{1}{2}
\sum_{\bf q}\overline{ |V_{{\bf q}11}-V_{{\bf q}22}|^2 }
\nonumber \\
&&\times\hbar^{-2} \int\nolimits_{-\infty}^{\infty}dt\left\langle \exp[i{\bf qr}_n(t)]\exp[-i{\bf qr}_n(0)]\right\rangle.
\end{eqnarray}

The frequency detuning (\ref{eq:Delta_omega}) depends on the distribution of SEs, cf. Eq.~(\ref{eq:11}), which makes the overall quantum kinetic equation (\ref{eq:12}) {\it nonlinear}. As shown below, this nonlinearity may result in the bistability of the response of the electron system to resonant radiation.

\subsection{Absorption bleaching}

To find the stationary distribution of the system, Eq.~(\ref{eq:12}) must be complemented with the equation that describes electron energy relaxation. However, an important conclusion about the distribution can be reached even without the analysis of energy relaxation in the case of weak to moderately strong microwave radiation, where
\begin{equation}
\label{eq:16}
\Omega_R^2\ll \gamma_0\gamma_{21}.
\end{equation}

Condition (\ref{eq:16}) is sufficient to suppress the conventional absorption saturation. Such saturation is well-known for two levels systems and requires that radiation makes the populations of the excited and ground states $\rho_{22}$ and $\rho_{11}$ close to each other, so that the probabilities of radiation-induced transitions up and down in energy become close, too. In the saturation regime the population ratio $\rho_{22}/\rho_{11}$ significantly differs from the Boltzmann factor $\exp[(\epsilon_1-\epsilon_2)/k_{\rm B}T_e]$. In our case instead of the levels we have subbands of electron motion. It is seen from Eq.~(\ref{eq:12}) that in the range (\ref{eq:16}) the deviation of $\rho_{22}/\rho_{11}$ from the Boltzmann factor is small even at resonance, $\delta\omega=0$, and even if one disregards transitions between subbands 1, 2 and other subbands. This shows that absorption saturation does not occur in our system. 

From Eq.~(\ref{eq:12}) and from the detailed balance condition (\ref{eq:14}) it follows that in the range (\ref{eq:16}) the overall stationary distribution of the electron system both over the energy of lateral motion and over the subbands of out-of-plane motion is characterized by the same temperature $T_e$, 
\begin{equation}
\label{eq:full_Boltzmann}
\rho_{\alpha\alpha}\approx Z^{-1}e^{-\epsilon_{\alpha}/k_{\rm B}T_e}, \qquad
Z=\sum_{\alpha}e^{-\epsilon_{\alpha}/k_{\rm B} T_e}.
\end{equation}

Even though there is no absorption saturation in the range (\ref{eq:16}), the absorption significantly changes with the increasing radiation intensity. The mechanism of this change is {\it absorption bleaching}.\cite{KonstantinovJPSJ2008,Ryvkin} It can be understood from Fig.~\ref{fig:1}. As a result of inter-subband scattering an electron that resonantly absorbed a photon in subband 1 and made a transition to subband 2 can go back to subband 1, but with higher energy. It can then again absorb a photon. Now it can be elastically scattered from subband 2  to subband 3. It can also go back to subband 1, again resonantly absorb a photon, and ne scattered into subbands 4, 5, etc. 

The cascade of photon-induced transitions and inter-subband scattering leads to population of higher and higher subbands and to a decrease of the population of subband 1. Such decrease, along with the fact that the populations of subbands 1 and 2 become closer with increasing $T_e$, lead to an overall decrease of resonant absorption.

\subsection{Bistability of resonant response }

A complete analysis of nonlinear response requires incorporating inelastic processes that lead to energy exchange between the electrons and excitations in helium. By now several types of inelastic processes have been identified, including the weak inelasticity of scattering by helium vapor atoms, the inelastic two-ripplon scattering and the inelastic scattering by phonons in liquid helium. For all of them, the energy relaxation rate depends on the electron state and helium temperature.\cite{Monarkha_book,Dykman2003,KonstantinovLTP2008,Schuster2010} 

The description of energy relaxation is simplified by the fact that it is slow. Between inelastic scattering events electrons have time to exchange energy with each other,  and the state of the electron system is described by electron temperature. Therefore one can introduce a single energy relaxation rate $\nu_E$, which is independent of the electron state and depends only on $T_e, T$, and $n_s$.\cite{Monarkha_book,Konstantinov2007,Monarkha2007,KonstantinovJPSJ2008,Ryvkin,KonstantinovLTP2008}. This leads to an implicit equation for $T_e$. It has the form of a simple energy balance equation
\begin{eqnarray}
\label{eq:17}
&& \nu_E k_{\rm B} (T_e-T) = \hbar\omega_F r Z^{-1} [e^{-\epsilon_1/k_{\rm B}T_e}-e^{-\epsilon_2/k_{\rm B}T_e}], \nonumber \\
&& r=\frac{1}{2}\Omega_R^2 \gamma_0/[\gamma_0^2 + (\omega_F - \omega_{21}-\Delta\omega_{21})^2].
\end{eqnarray}
We have introduced here absorption coefficient $r$, which is determined by Eqs.~(\ref{eq:12}) and (\ref{eq:Delta_omega}). The left- and right-hand sides of Eq.~(\ref{eq:17}) are, respectively, the energies dissipated into helium and absorbed by an electron from radiation per unit time. For all energy relaxation mechanisms considered so far, the left-hand side of Eq.~(\ref{eq:17}) is an increasing function of $T_e-T$, at least for not too large $T_e - T$.

\begin{figure}[h]
\includegraphics[width=7.5cm]{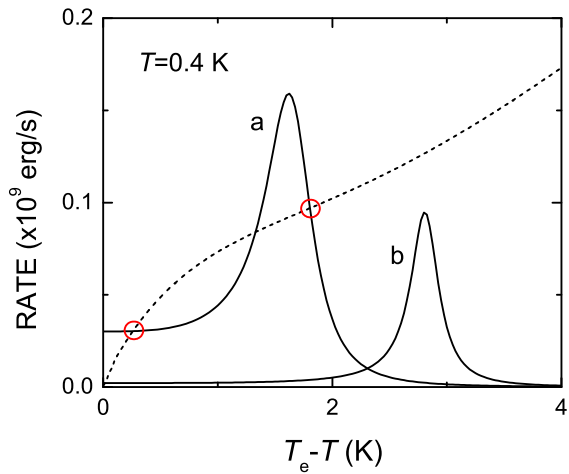}
\caption{(color online) Graphical solution of the energy balance equation (\ref{eq:17}) for $T=0.4$~K and $n_s=4.0\times 10^7$~cm$^2$. The dashed line is the energy loss rate [the left hand side of Eq.~(\ref{eq:17})] due to inelastic scattering by helium vapor atoms. The solid lines show the rate of the microwave power absorption [the right hand side of Eq.~(\ref{eq:17})] calculated for $\omega_F=104.5$~GHz, $\Omega_R=2.0$~MHz,  and  $\omega_F-\omega_{21}=0.1$~GHz (a) and 0.4~GHz (b). For the case (a), the stable solutions of Eq.~(\ref{eq:17}), which are given by the intersections of solid and dashed lines, are marked by (red) circles.}
\label{fig:2}
\end{figure}

Parameter $r$ is a nonlinear function of $T_e$, as seen from Eqs.~(\ref{eq:11}) and (\ref{eq:Delta_omega}). Therefore  Eq.~(\ref{eq:17}) is fairly complicated. A convenient way of understanding the possible types of its solution is based on the graphical solution, an example of which is shown in Fig.~\ref{fig:2}. The plot shows the rates of energy absorption and relaxation calculated for $T=0.4$~K where the scattering is dominated by interaction with helium vapor atoms. In this regime, the $T_e$-dependence of $\nu_E$, $\gamma_0$ and $\Delta\omega_{21}$ are calculated for the typical experimental parameters with account taken of 200 subbands of electron motion normal to the helium surface. 

The left-hand side of Eq.~(\ref{eq:17}) is shown in Fig.~\ref{fig:2} by a dashed line, while the right-hand side (rhs) is shown by the solid lines for $\omega_F-\omega_{21}=0.1$ and 0.4~GHz (lines a and b, respectively). The major part of $T_e$-dependence in the rhs comes from the frequency shift $\Delta\omega_{21}$. From Eq.~(\ref{eq:11}), $\Delta\omega_{21}$ rather quickly increases with the increasing $T_e$. In a certain range of $\omega_F-\omega_{21}$, the change of $T_e$ can bring $\Delta\omega_{21}$ in resonance with $\omega_F-\omega_{21}$. This leads to a resonant peak of $r$ and thus of the rhs of  Eq.~(\ref{eq:17}) as function of $T_e$ seen in Fig .~\ref{fig:2}. 

The heights of the resonant peaks in Fig.~\ref{fig:2} depend on $\omega_F-\omega_{21}$. From Eq.~(\ref{eq:17}), the coefficient in front of $r$ decreases with increasing $T_e$. The maximal value of $r$ at resonance, $r_{\max}\propto 1/\gamma_0$, also decreases. Therefore the higher is the value of $T_e$ where $\Delta\omega_{21}=\omega_F-\omega_{21}$, the smaller is the height of the peak of the rhs of Eq.~(\ref{eq:17}). 

The solutions of Eq.~(\ref{eq:17}) are determined by the crossings of the dashed and solid line in Fig.~\ref{fig:2}. They give the effective temperature $T_e$ in the stationary state. Where the microwave frequency is relatively far from resonance, i.e., $\omega_F-\omega_{21}$ is relatively large, there exists only one solution for $T_e$, with $T_e$ close to the helium temperature $T$. Closer to resonance, on the other hand, Eq.~(\ref{eq:17}) can have two stable solutions with comparatively large and small $T_e-T$. They are marked by circles in Fig.~\ref{fig:2}. In this case Eq.~(\ref{eq:17}) also has a  stationary solution for intermediate $T_e$, as seen from Fig.~\ref{fig:2}. By writing the energy change per unit time as the difference of the right- and  left-hand sides of Eq.~(\ref{eq:17}), one can see that this intermediate solution is unstable. 

The left and right hand sides of Eq.~(\ref{eq:17}) display a qualitatively similar behavior for other parameters and for other relaxation mechanisms. This indicates that the onset of bistability is a general consequence of absorption bleaching.

The occurrence of two stable regimes can be understood in the following way. In one regime, which corresponds to lower $T_e$, the overall detuning of the electron transition frequency $\omega_{21}+\Delta\omega_{21}$ from the radiation frequency $\omega_F$ is large compared to $\gamma_0$. Therefore, the absorption rate is comparatively small and $T_e$ is self-consistently low. On the other hand, if $T_e$ is sufficiently high, $\omega_{21}+\Delta\omega_{21}$ is close to $\omega_F$, absorption is comparatively strong and $T_e$ is self-consistently high. We note that, since $\Delta\omega_{21}$ is positive, such bistability can occur only for $\omega_F-\omega_{21}>0$.

The bistability leads to the hysteresis of electron absorption with varying $\omega_F - \omega_{21}$. If we start from large detuning $\omega_F - \omega_{21}$ (curve $A$ in Fig.~\ref{fig:2}) and then decrease it, the system will stay in the small-absorption state, which corresponds to left circle in Fig.~\ref{fig:2}. On the other hand, if we start from small $\omega_F - \omega_{21}$, the system will be on the decreasing with increasing $T_e$ part of the resonant solid line in Fig.~\ref{fig:2}. As $\omega_F - \omega_{21}$ increases, the system will get into the region of strong absorption, but ultimately will switch to the low-absorption branch for large $\omega_F - \omega_{21}$. Below we present the experimental observation of frequency shift, bistability, and hysteresis and make a comparison with the theory.

\section{Experiment}

The details of the experimental apparatus have been described elsewhere \cite{KonstantinovLTP2008}. SEs were accumulated on the surface of liquid $^3$He placed about half way between two round electrodes of diameter 26~mm separated by approximately 2.6~mm and forming a parallel-plate capacitor. A positive voltage $V_{\rm B}$ was applied to the bottom electrode to create an electric field $E_{\perp}$ perpendicular to the helium surface. In addition, a magnetic field $B$ was applied perpendicular to the  surface using a superconducting magnet placed around the experimental cell. Microwave radiation at fixed frequency $\omega_F/2\pi=104.5$~GHz was passed through the experimental cell, and the inter-subband splitting $\epsilon_2-\epsilon_1$ was brought in resonance with $\hbar\omega_F$ by varying $E_{\perp}$.

The top plate of the capacitor consisted of two concentric-ring electrodes (the Corbino disk geometry), which were used to measure the longitudinal magneto-conductivity of the electron system $\sigma_{xx}$.  The Corbino signal proportional to $\sigma_{xx}^{-1}$ was recorded while slowly varying $V_{\rm B}$ at a fixed microwave power and at a fixed value of the magnetic field $B$. The variation of $\sigma_{xx}$ caused by the radiation-induced heating of SEs allowed us to observe resonant response from inter-subband absorption. Subtracting the value of the signal recorded in the absence of radiation, the relative change $\Delta\sigma_{xx}^{-1}/\sigma_{xx}^{-1}$ due to the heating was determined and plotted vs $V_{\rm B}$. The measurements reported here were done in the temperature range from 0.2 to 0.4~K.

For each curve $\Delta\sigma_{xx}^{-1}/\sigma_{xx}^{-1}$ vs $V_{\rm B}$, the electron temperature $T_e$ was calculated from the previously discussed dependence on $T_e$ of the quasi-elastic electron transport time $\tau$.\cite{Konstantinov2007,KonstantinovLTP2008} For magnetic fields $B$ used in the experiment, the many-electron conductivity $\sigma_{xx}$ follows the Drude law,\cite{Dykman1993}
\begin{equation}
\sigma_{xx}^{-1}=(1+\mu^2 B^2)/\sigma_0,
\label{eq:Drude}
\end{equation}
where $\sigma_0=n_s e\mu$ is the conductivity in zero magnetic field and $\mu=e\tau/m$ is the electron mobility. For $\mu B \gg 1$, $\sigma_{xx}^{-1}$ is proportional to $\tau$. To find the proportionality coefficient, which depends on the electron density $n_s$, the Corbino signal was recorded without radiation while SEs were slowly cooled down until they formed a Wigner crystal. The crystallization could be easily detected as an abrupt change in $\sigma_{xx}$.\cite{Shirahama1995} The value of  $n_s$ was found from the crystallization temperature and the well established critical value of the plasma parameter $\Gamma=e^2(\pi n_s)^{1/2}/k_BT\approx 130$.

The experimental procedure for direct absorption measurements was similar to that described in Ref.~\onlinecite{Isshiki2007}. The power of the microwave radiation that passed through the cell was measured with an InSb bolometer mounted inside the cryostat. Resonant absorption could be observed as the variation of the bolometer signal at $T=0.4$~K as electrons were driven through resonance by sweeping the bottom plate voltage $V_{\rm B}$. Simultaneously, this voltage was modulated with a small sinusoidal signal at frequency 10~kHz, and the in-phase and quadrature components of the demodulated bolometer signal were obtained using a dual-phase lock-in amplifier. With such an arrangement, in the limit of small modulation amplitude the  in-phase component of the obtained signal is proportional to the derivative of the power absorbed by SEs with respect to $V_{\rm B}$. Normally, the quadrature component of the signal should be zero. The absorption lineshape was obtained by numerical integration of the signal with respect to $V_{\rm B}$.

\section{Results}

\subsection{Resistivity measurements}

\begin{figure}[h]
\includegraphics[width=7.5cm]{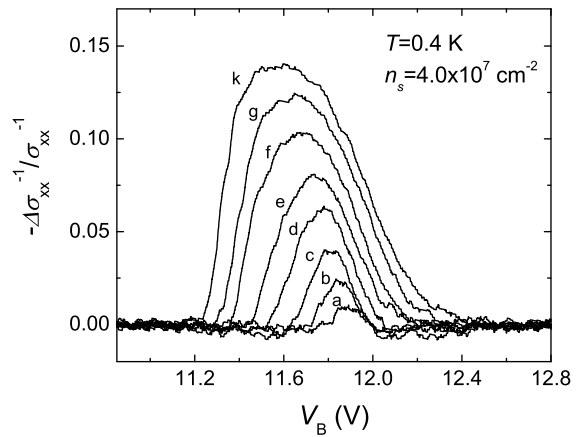}
\caption{Relative change of the reciprocal conductivity $-\Delta\sigma_{xx}^{-1}/\sigma_{xx}^{-1}$ vs the voltage applied to the bottom electrode $V_{\rm B}$ obtained for SEs on liquid $^3$He at $T=0.4$~K, $n_s=4\times 10^7$~cm$^{-2}$ and several values of the input power of radiation. The lines $a$ to $k$ correspond to -17.5, -14.0, -11.4, -8.8, -6.2, -4.2, -2.3 and -0.6 dB of the input power expressed as a fraction (in decibel) of the maximum power $\sim 1$~mW.}
\label{fig:shift}
\end{figure}

In Fig.~\ref{fig:shift} we show the relative radiation-induced conductivity change $-\Delta\sigma_{xx}^{-1}/\sigma_{xx}^{-1}$ as a function of the voltage that presses the electrons to the helium surface $V_{\rm B}$. All plots were obtained from the data taken at $T=0.4$~K and electron density $n_s=4\times 10^7$~cm$^{-2}$. The radiation power was measured at the output of the microwave source;  shown in Fig.~\ref{fig:shift} is the ratio of the applied power (in decibel) to the maximum power, which was approximately 1 mW. Experimental traces were obtained by increasing $V_{\rm B}$ slowly to drive SEs through resonance. A conversion of the increase of $V_{\rm B}$  to the increase of the transition frequency $\omega_{21}$ by $\approx 25$~MHz is discussed in Sec.~V~A.

At $T=0.4$~K, electron scattering is mostly determined by collisions with helium vapor atoms. As we showed previously,\cite{Konstantinov2007,KonstantinovLTP2008} in this case radiation-induced heating leads to a decrease of the relaxation time $\tau$. Therefore $\sigma_{xx}^{-1}$ should decrease at resonance. 

The resistivity curves in Fig.~\ref{fig:shift} show a pronounced resonant peak. The height of the peak monotonically increases with the input power. At small power levels, the resistivity shows a slight increase on the sides of the resonance (respectively, $-\Delta\sigma_{xx}^{-1}$ becomes negative). This effect is attributed to the contribution of ripplon scattering to the momentum relaxation rate.\cite{KonstantinovLTP2008} As the power increases, the resonance shifts toward lower values of $V_{\rm B}$, which corresponds to the increase in the transition frequency. Also, the lineshape becomes asymmetric. We believe the shift is due to the many-electron effect described by Eq.~~(\ref{eq:11}) and is associated with thermal population of the excited subbands as the effective electron temperature increases.

\begin{figure}[h]
\includegraphics[width=7.5cm]{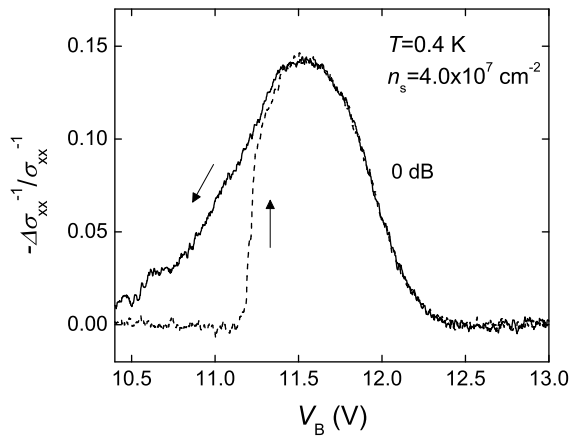}
\caption{Relative change of the reciprocal conductivity $-\Delta\sigma_{xx}^{-1}/\sigma_{xx}^{-1}$ vs $V_{\rm B}$ obtained at $T=0.4$~K, $n_s=4\times 10^7$~cm$^{-2}$ and at the maximum input power (0dB). The dashed and solid lines correspond to sweeping through the resonance from the low and high side of $V_{\rm B}$, respectively, as indicated by the arrows.}
\label{fig:HThyst}
\end{figure}

At high power levels, when the voltage is swept through the resonance from the low-$V_{\rm B}$ side, the resistivity jumps abruptly to a lower value; respectively, $-\Delta\sigma_{xx}^{-1}/\sigma_{xx}^{-1}$ jumps up. This unusual behavior is demonstrated in Fig.~\ref{fig:HThyst}. The data used for this plot were taken at similar conditions to those in Fig.~\ref{fig:shift}, except that the sweeping rate was slower by a factor of two. The jump always occurs on the low-field side of the resonance and is well reproducible. When the sweeping direction is reversed, the resistivity curve first follows the forward-sweeping curve as it passes through the resonance. However, on the low-field side it does not show any jump but slowly goes to zero with decreasing $V_{\rm B}$. This behavior is also well reproducible. Therefore, overall the system displays a well-pronounced hysteresis.

\begin{figure}[h]
\includegraphics[width=7.5cm]{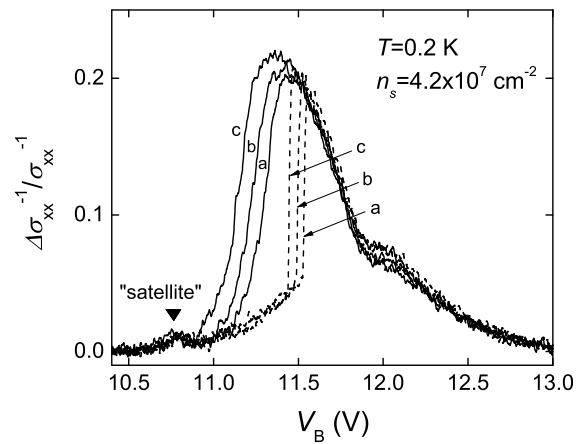}
\caption{Relative change of the reciprocal conductivity $\Delta\sigma_{xx}^{-1}/\sigma_{xx}^{-1}$ vs $V_{\rm B}$ obtained at $T=0.2$~K, $n_s=4.2\times 10^7$~cm$^{-2}$ and three different values of the input power: -6.2, -5.4 and -4.2~dB (lines a, b and c, respectively). The dashed and solid lines correspond to the sweeping through the resonance from the low- and high-$V_{\rm B}$ sides, respectively. A satellite peak that appears on the low-field side of the resonance is indicated by the full triangle.}
\label{fig:LThyst}
\end{figure}

The resistivity jump becomes more prominent and the width of the hysteresis loop increases at lower temperatures. In Fig.~\ref{fig:LThyst} we plot $\Delta\sigma_{xx}^{-1}/\sigma_{xx}^{-1}$ vs $V_{\rm B}$ for the data taken at $T=0.2$~K, $n_s=4.2\times 10^7$~cm$^{-2}$, and at three different input power levels. At this temperature, the relaxation rate is determined by the interaction of the electrons with ripplons, and electron heating leads to an increase of the relaxation time $\tau$. Therefore $\sigma_{xx}^{-1}$ increases at resonance. \cite{KonstantinovLTP2008} In addition to the abrupt jump and hysteresis observed at $T=0.4$~K, the lineshape shows an interesting structure. It seems to be a superposition of a broad peak centered at $V_{\rm B}\approx 11.9$~V and a high peak located on the low-$V_{\rm B}$ side of the broad peak. While this latter peak grows and shifts with the increasing input power, the position of the broader peak does not change and the height of the peak slightly decreases with the increasing power. We identify the high peak as being due to resonant transitions from the ground state to the first excited state. The broader peak remains unexplained at this time. 

In addition to the peaks described above, a small satellite peak appears on the low-field side of the resonance. This peak becomes visible only at very high power levels. It was reported in our previous work~\cite{KonstantinovJLTP2007} and was attributed to the excitation of collective in-plane electron vibrations.\cite{PlatzmanTzoar1976} At extremely high radiation intensities, we observe a number of satellite peaks on the both sides of the resonance. Some of them can be due to resonant transitions between higher excited subbands and emerge through self-sustained absorption, which was recently discovered in our experiments.\cite{KonstantinovJLTP2010} The origin of other peaks is not clear at the moment. A detailed description of the satellite structure of the resonance will be given elsewhere.
\begin{figure}[h]
\includegraphics[width=7.5cm]{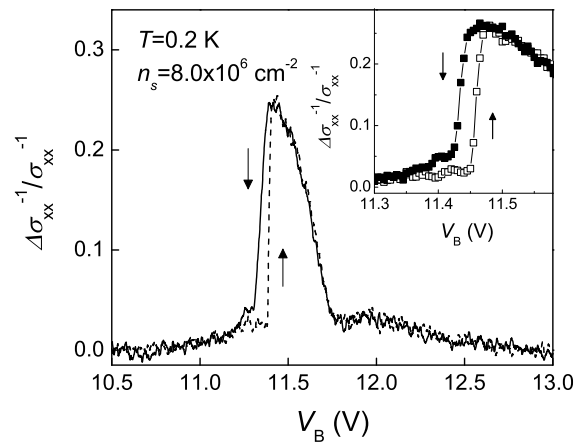}
\caption{$\Delta\sigma_{xx}^{-1}/\sigma_{xx}^{-1}$ vs $V_{\rm B}$ for $T=0.2$~K, $n_s=8.0\times 10^6$~cm$^{-2}$, and -2.3~dB of the input power. The dashed and solid lines correspond to the sweeping through the resonance from the low and high $V_{\rm B}$ side, respectively, as indicated by arrows. Inset: plots obtained under the same conditions, but with the slower by a factor of two rate of sweeping from the lower (opened squares) and the higher (solid squares) $V_{\rm B}$ side.}
\label{fig:tooth}
\end{figure}

The signal strongly changes at lower electron density. In Fig.~\ref{fig:tooth} we plot $\Delta\sigma_{xx}^{-1}/\sigma_{xx}^{-1}$ vs $V_{\rm B}$ for the data taken at the same  temperature as in Fig.~\ref{fig:LThyst} but at $n_s=8.0\times 10^6$~cm$^{-2}$. Here, the intensity (area) of the broad peak is significantly smaller, while the higher peak is much narrower. Interestingly, for lower electron density the signal shows abrupt jumps for both sweeping directions. It has a characteristic sawtooth shape expected for switching between two stable states.The abrupt switching for sweeping the voltage up and down is highlighted in the inset of Fig.~\ref{fig:tooth}.

\subsection{Absorption measurements}

We have also studied absorption directly by measuring the power of the microwave radiation passed through the electron system. The experiment was done by sinusoidally modulating the potential $V_{\rm B}$ and sweeping the central point $\bar V_{\rm B}$ about which the modulation was performed. The measured transmitted power was then Fourier-transformed using a lock-in amplifier. Unexpectedly we found that, in the absorbed power, along with the standard in-phase component $I(\bar V_{\rm B})$, for high radiation power the signal displayed a quadrature component $Q(\bar V_{\rm B})$. The data were obtained for the modulation frequency $\omega_m/2\pi$=10 kHz and the modulation amplitude $V_m= 10$~mV, which corresponds to the modulation amplitude $\approx 25$~MHz of the transition frequency $\omega_{21}$ (see below). For modulation $\propto \cos\omega_mt$ we define, following the standard notations, the quadrature component as a coefficient at $-\sin\omega_mt$.

\begin{figure}
\includegraphics[width=7.5cm]{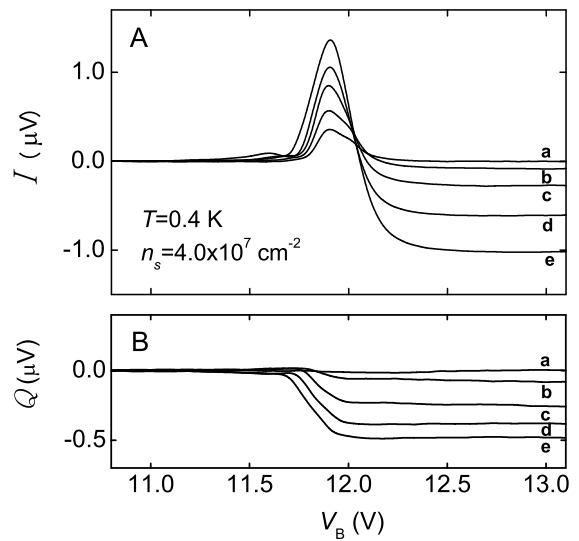}
\caption{Integrated lineshape of the absorption signal measured with SEs on liquid $^3$He for $T=0.4$~K, $n_s=4\times 10^7$~cm$^{-2}$ at different power levels. The graph shows the in-phase (panel A) and the quadrature (panel B) components of the measured signal. The lines $a$ to $e$ correspond to -27.7, -20.6, -17.5, -14.4, and -11.4~dB  of the input power attenuation.}
\label{fig:hyst}
\end{figure}

In Fig.~\ref{fig:hyst} we show functions
\begin{equation}
\label{eq:cal_I_Q}
{\cal I}(V_{\rm B})=\int^{V_{\rm B}}dV_{\rm B}' I(V_{\rm B}'),\qquad {\cal Q}(V_{\rm B})=\int^{V_{\rm B}}dV_{\rm B}'Q(V_{\rm B}'),
\end{equation}
sometimes called integrated lineshapes. They were obtained by sweeping $\bar V_{\rm B}$ upward for $T=0.4$~K and $n_s=4\times 10^7$~cm$^{-2}$ at different power levels. The lower limits of the integrals in Eq.~(\ref{eq:cal_I_Q})  were chosen at a value of $\bar V_{\rm B}$ well below the resonance; the result weakly depended on this limit, as it is clear from Fig.~\ref{fig:hyst}. For $T=0.4~K$ the intrinsic linewidth of the absorption spectrum is rather small and the observed lineshape is determined by the inhomogeneous broadening due to the nonuniformity of the  electric field at the helium surface as well as the nonuniformity of the microwave field~\cite{Glasson2004,Isshiki2007}. 

As expected, for low radiation power the quadrature component of the absorption signal is zero. However,
at higher powers hysteresis occurs within each modulation cycle itself. This gives a distinctive quadrature component in the modulated output, see Sec.~V. The effective lineshape which is obtained from the integrated modulation signal then has a quadrature component and an offset in both components. Note that the quadrature component itself is not hysteretic - it is the same sweeping the voltage up or down. But this component gives a sensitive indication of local hysteresis in each modulation cycle. The features of the integrated lineshape become more prominent with the increasing power. For lower $T$, the quadrature component is seen for lower power levels.

\section{Discussion}

\subsection{The many-electron frequency shift}

The resonance condition for exciting an electron from the ground to the first excited level of motion normal to the helium surface is determined by the many-electron frequency shift, $\omega_F-\omega_{21}=\Delta\omega_{21}$ with $\Delta\omega_{21}(T_e)$ given by Eq.~\ref{eq:11}. The position of the resonance depends on the electron temperature $T_e$. In turn, $T_e$ depends on the absorbed power and for fixed incident power is expected to be maximal at resonance.  In the parameter ranges used in our experiment $\sigma_{xx}^{-1}$ varies monotonically with $T_e$. Therefore we assume that the maxima of the resistivity curves plotted in Fig.~\ref{fig:shift} correspond to exact resonance. Then, for each power level, the frequency shift $\Delta\omega_{21}$ can be found from the shift of the maximum of the corresponding curve with respect to the resonant absorption curve measured at very low power using the InSb bolometer. For this curve one can assume $T_e=T$, and then in Eq.~(\ref{eq:11}) for $\Delta\omega_{21}$  one can set $\rho_{\alpha\alpha}=\delta_{\alpha,1}$. Generally, the parameters $z_{\alpha\beta}$ in Eq.~(\ref{eq:11}) depend on the pressing field $E_{\perp}$ and thus on $V_{\rm B}$; however, this dependence is smooth, and in the narrow range of $V_{\rm B}$ that we studied it could be disregarded. 

To find the shift in frequency units, the frequency of the microwave source was changed by a known amount and the corresponding shift in $V_{\rm B}$ was recorded. Such measurements were done at low power  so that the many-electron shift remained constant. This procedure allowed us to establish the relationship between $\omega_{21}$ and $V_{\rm B}$. The conversion factor, i.e., the slope of $\omega_{21}/2\pi$ vs $V_{\rm B}$, was found to be $2.5\pm 0.1$ GHz/V.
\begin{figure}[h]
\includegraphics[width=7.5cm]{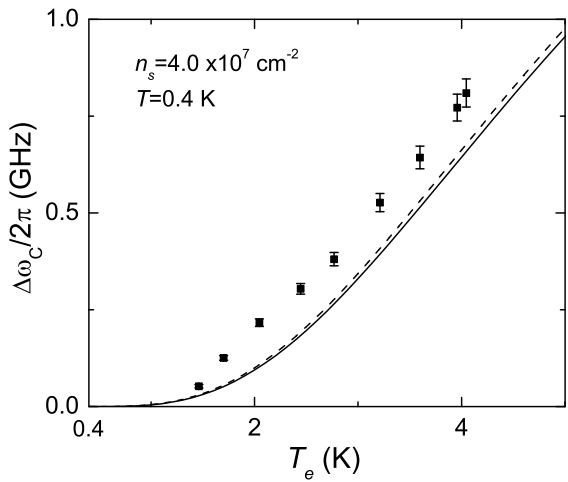}
\caption{Many-electron frequency shift $\Delta\omega_{21}$ vs electron temperature $T_e$ obtained from the data shown in Fig.~\ref{fig:shift} (squares); $\Delta\omega_{21}$ is counted off from its value for $T_e=T=0.4$~K . The solid line shows the result of calculations using Eq.~\ref{eq:11} for $n_s=4\times 10^7$~cm$^{-2}$. The dashed line shows the calculation that excludes the term proportional to $|z_{12}|^2$ in Eq~(\ref{eq:11}).}
\label{fig:compar}
\end{figure}

In Fig.~\ref{fig:compar} we plot the many-electron frequency shift and the electron temperature determined for the maximum of each curve in Fig.~\ref{fig:shift}. We remind that $T_e$ was obtained by comparing the measured $\sigma_{xx}^{-1}$ with its theoretical value for given $n_s$ and $T$. The frequency shift is counted off from its value for $T_e=T$. It monotonically increases with $T_e$ and reaches the value of about 1~GHz for the highest-power curve in Fig.~\ref{fig:shift}. For comparison, we also show in Fig.~\ref{fig:compar} the frequency shift calculated from Eq.~\ref{eq:11} for $n_e=4\times 10^7$~cm$^{-2}$. In this calculation we limited the number of the states of out of plane motion to 20, which was sufficient for the studied range of $T_e$. At high $T_e$, the shift observed in the experiment is about 20$\%$ larger than our theoretical estimate. This discrepancy might be attributed to the nonuniformity of the electron density and the microwave power across the electron layer and also to the corrections to the approximations used to obtain Eq.~\ref{eq:11} and to evaluate $T_e$.

\subsection{Nonlinear absorption and bistability}

The experimental results on the bistability and hysteresis of nonlinear response discussed in Sec.~IV (see Figs.~\ref{fig:HThyst}-\ref{fig:tooth}) are in qualitative agreement with the theory presented in Sec.~II. The data clearly demonstrate hysteresis: the state of the system depends on the direction of sweeping $V_{\rm B}$ or, equivalently, the single-electron transition frequency  $\omega_{21}$. The abrupt jump in $\Delta\sigma_{xx}^{-1}/\sigma_{xx}^{-1}$ observed in the experiment is due to switching between different states of the electron system that coexist in a limited parameter range. Switching occurs once the corresponding state disappears with the varying parameter. 

Sharp jumps in Figs.~\ref{fig:HThyst} - \ref{fig:tooth} were observed with increasing $\omega_{21}$ for switching from the states with comparatively small absorption and low $T_e$ to high-$T_e$ states. Such switching is expected from the theory and can be inferred from Fig.~\ref{fig:2}. If one thinks of moving curve B in Fig.~\ref{fig:2} to the left, one can see that the low-$T_e$ state will disappear. For data taken at $T=0.4$~K (see Fig.~\ref{fig:HThyst}), $T_e$ jumps from about 0.4~K to about 3.1~K. For data taken at $T=0.2$~K and with similar electron density (see Fig.~\ref{fig:LThyst}), the jump is from about 0.3~K to about 2.6~K. 

For the reversed sweeping direction, from higher to lower $\omega_{21}$, the response displays an abrupt jump only for low electron densities, see Fig.~\ref{fig:tooth}. For higher densities a sharp switching to the small-absorption state was not observed, instead the response smoothly varied with the parameters in the range of hysteresis.  This indicates that the dynamics of the system is more complicated than the model of Sec.~II suggests. In particular, the spatial inhomogeneity of the electron density and the microwave field should play a role. 

The inhomogeneity would favor a sharper upward jump of the absorption. One can think that, with the increasing $V_{\rm B}$, there is formed a ``critical nucleus" of high absorption and high $T_e$, which then quickly expands throughout the system. On the other hand, with the decreasing $V_{\rm B}$, the areas that exist only in the low $T_e$-state gradually  increase in size. Regrettably, these processes are hard to characterize quantitatively without a detailed picture of the radiation power distribution. Another indication of a more complicated nature of the system is the broad peak of $\Delta\sigma_{xx}^{-1}/\sigma_{xx}^{-1}$ vs $V_{\rm B}$ in Figs.~\ref{fig:LThyst} and \ref{fig:tooth}, which the model of Sec.~II does not explain.  We note that the narrowing of the hysteresis loop for lower density in Fig.~\ref{fig:tooth} compared to Fig.~\ref{fig:LThyst} can be understood by noticing that $\Delta\omega_{21}$ scales with the electron density as $n_s^{3/2}$ for the same $T_e$. This gives a factor of 12 difference between the values of $\Delta\omega_{21}$ for the densities in Figs.~\ref{fig:LThyst} and \ref{fig:tooth}.

\subsection{Absorption signal for modulation across the hysteresis loop}

The theory of Sec.~II also explains the complicated behavior of the microwave power absorption described in Sec.~IV.~B, see Fig.~\ref{fig:hyst}. This type of behavior was also reported by Glasson {\it et al.} for SEs on liquid $^4$He and was attributed to hysteresis of microwave absorption.\cite{Glasson2004} In the absorption experiment, the transition frequency $\omega_{21}(t)$ is swept up and down with a modulation period $T_0$. Within a modulation period, the absorption for increasing and decreasing $\omega_{21}(t)$ can be different. This happens if the modulation amplitude is large enough, so that $\omega_{21}(t)$ goes across the whole region where the two stable states of the electron system coexist, i.e., across the hysteresis region.

We now show that hysteresis leads to the behavior seen in Fig.~\ref{fig:hyst}. Suppose the frequency is modulated sinusoidally, $\omega_{21}(t)-\bar\omega_{21}=A \cos(\omega_m t)$, where the overline means period average. In the experiment, $\bar\omega_{21}$ is determined by the value of the potential $\bar V_{\rm B}$ about which the modulation is performed, $\bar\omega_{21}\equiv \bar\omega_{21}(\bar V_{\rm B})$. For the chosen modulation phase, $\omega_{21}(t)$ is swept down during the first half of the modulation period $\pi/\omega_m$ and up during the second half. 

If $\omega_{21}(t)$  goes across the hysteresis region, the absorbed microwave power $S(\omega_{21}(t))$ takes on  different values, $S_-(\omega_{21}(t)$) and $S_+(\omega_{21}(t))$ on the down and up sweeps, respectively; note that functions $S_{\pm}$ have discontinuities where the absorption switches between the high- and low-absorption branches, which we denote by $S_>(\omega_{21})$ and $S_<(\omega_{21})$, respectively. Where $\omega_{21}(t)$ goes across the hysteresis region, function $S_-$ first evolves along the branch $S_>$ and then switches to $S_<$, whereas function $S_+$ evolves along $S_<$ and then switches to $S_>$.

The in-phase and quadrature components of the absorbed power detected with a lock-in amplifier, $I\equiv I(\bar V_{\rm B})$ and $Q\equiv Q(\bar V_{\rm B})$, are
\begin{eqnarray}
I=\frac{\omega_m}{\pi}\int\limits_0^{\pi/\omega_m}\cos(\omega_mt)\left[S_-(\omega_{21}(t))+S_+(\omega_{21}(t))\right]dt, \nonumber \\
Q=\frac{\omega_m}{\pi}\int\limits_0^{\pi/\omega_m}\sin(\omega_mt)\left[S_+(\omega_{21}(t))-S_-(\omega_{21}(t))\right]dt.
\label{eq:fourier}
\end{eqnarray}

From Eq.~(\ref{eq:fourier}), where the modulation cycle goes across the hysteresis loop the signal has a nonzero quadrature component,
\begin{equation}
Q=\frac{1}{\pi A}\int\nolimits_{-A}^A \left[ S_+(\bar\omega_{21}+x)-S_-(\bar\omega_{21}+x) \right]dx.
\label{eq:imag}
\end{equation}
The value of $Q$ does not depend on what part of the modulation cycle is covered beyond the hysteresis region, as outside this region $S_+=S_-$.

We now explain the behavior of the experimentally measured parameters ${\cal I}$ and ${\cal Q}$ in Fig.~\ref{fig:hyst}  where, as $\bar \omega_{12}$ is increased, it goes through the hysteresis region in such a way that the system switches from the lower- to a higher-absorption branch. We start with the quadrature component ${\cal Q}$. 
Let us assume that hysteresis occurs in the region $\omega_L < \omega_{21} <\omega_H$. If the modulation amplitude $A < (\omega_H-\omega_L)/2$ then as the voltage $\bar V_{\rm B}$ is swept up, the absorption will instantaneously switch from the low to the high branch. In this case the quadrature component $Q = 0$. But if $A > (\omega_H-\omega_L)/2$, the hysteresis loop will occur in each modulation cycle, in the appropriate range of $\bar\omega_{21}(\bar V_{\rm B})$. Then
\begin{eqnarray}
\label{eq:Q_within_hysteresis}
Q\equiv Q(\bar V_{\rm B})=(\pi A)^{-1}\int_{\omega_L}^{\omega_H}dx[S_<(x)-S_>(x)]\nonumber\\
\times\left[ \theta(\bar\omega_{21}-\omega_H +A) - \theta(\bar\omega_{21}-\omega_L - A) \right],
\end{eqnarray}
where $\theta(x)$ is the Heaviside step function and $\bar\omega_{21}\equiv \bar\omega_{21}(\bar V_{\rm B})$; we used that $S_+(\omega)=S_<(\omega)$ and $S_-(\omega)=S_>(\omega)$ for $\omega_L < \omega < \omega_H$.

From Eq.~(\ref{eq:Q_within_hysteresis}), $Q(\bar V_{\rm B})$ remains constant for $\omega_H-A <\bar \omega_{21} <\omega_L+A$. This constant is negative and is equal to $-S_{\square}/\pi A$, where  $S_{\square}$ is the area of the  hysteresis loop on the plane $(\omega_{21}, S)$. Respectively, the integrated quadrature ${\cal Q}(\bar V_{\rm B})$ linearly decreases with increasing $\bar V_{\rm B}$ in this range and becomes constant once $\bar\omega_{21}(\bar V_{\rm B})$ reaches $\omega_L+A$, in agreement with Fig.~\ref{fig:hyst}. This constant, which is the offset of the  integrated quadrature $\Delta{\cal Q}$ on the high-$V_{\rm B}$ side, is given by
\begin{equation}
\label{eq:Q_offset}
\Delta{\cal Q}= -(2C_V/\pi)S_{\square}\left[1-(\omega_H-\omega_L)/2A\right], 
\end{equation}
where $C_V=(d\omega_{21}/dV_{\rm B})^{-1}$ is the reciprocal slope of transition frequency vs voltage. 

The offset of the integrated quadrature is a clear indication of hysteresis in the system, which thus can be revealed by studying the response for a finite modulation amplitude. The increase of the slope of the curves in Fig.~\ref{fig:hyst} with increasing power corresponds to the increase of the area of the hysteresis loop $S_{\square}$.

We now consider the in-phase component. If function $S(\omega)$ were single-valued and smooth, $S(\omega)=S_-(\omega)=S_+(\omega)$, in the limit of small modulation amplitude $A\to 0$ we would have $I(V_{\rm B})\approx A\partial_{\omega}S$, which gives ${\cal I}(V_{\rm B})\approx AC_V S\bigl(\omega_{21}(V_{\rm B})\bigr)$. Then ${\cal I}(V_{\rm B})$ goes to zero on the both sides of the absorption peak.

The behavior of ${\cal I}$ becomes different in the presence of the hysteresis loop. We assume that the direction of sweeping $V_{\rm B}$ is such that we move along the low-absorption branch and jump to the high-absorption branch, that is, in our system, we increase $V_{\rm B}$.  In the limit $A\to 0$ (in particular, $A\ll \omega_H-\omega_L$), we have ${\cal I}(V_{\rm B})\approx AC_V S_<\bigl(\omega_{21}(V_{\rm V})\bigr)$ on the low-absorption branch $S_<(\omega_{21})$, i.e., for $\omega_{21}<\omega_H$. After the jump to the high-absorption branch $S_>(\omega_{21})$ we have, as before, $I(V_{\rm B})\approx A\left[\partial_{\omega}S_>\right]_{\omega_{21}}$. Then, for $\omega_{21}>\omega_H$, ${\cal I}(V_{\rm B})\approx AC_V \left[S_>\bigl(\omega_{21}(V_{\rm B})\bigr) -  \Delta S\right]$, where $\Delta S=S_>(\omega_H) - S_<(\omega_H-0)$ is the height of the jump of $S$ at $\omega_H$.
This shows that, as $V_{\rm B}$ will have gone over the resonant absorption peak, where $S_>\bigl(\omega_{21}(V_{\rm B})\bigr)\to 0$,  ${\cal I}$ will become negative. 

The offset $\Delta{\cal I}$ on the high-$V_{\rm B}$ side is given by the height of the jump of the absorption coefficient, $\Delta{\cal I}=-AC_V\Delta S$.
Such offset is a characteristic feature of the hysteresis. It sensitively depends on the interrelation between the modulation amplitude and the width of the hysteresis loop.\cite{Glasson2004}

 We now discuss the case where the modulation range $2A$ exceeds the width of the hysteresis region and assume that this region is narrow. To the leading order in $\omega_H-\omega_L$ and in $A$, for $\omega_L+A > \bar\omega_{21} > \omega_H-A$ we have $I(\bar V_{\rm B}) \approx \pi^{-1}\Delta S [\sin\omega_mt_L + \sin\omega_m t_H]$, where the values of $t_{L,H}$ are given by equations $\bar\omega_{21} + A\cos\omega_mt_{L,H}=\omega_{L,H}$ with $0 <\omega_mt_{L,H} < \pi$. This order of magnitude estimate is obtained by disregarding the change of $S_>,S_<$ within the hysteresis loop.

The positive value of $I$ reduces the negative offset of ${\cal I}$. Integrating  the above expression for $I(\bar V_{\rm B})$ over $\bar V_{\rm B}$ in the range where $\omega_L+A > \bar\omega_{21}(\bar V_{\rm B})  > \omega_H-A$, we obtain for the offset
\begin{eqnarray}
\label{eq:offset}
&&\Delta{\cal I}\approx \pi^{-1}AC_V\Delta S\left[\aleph(1-\aleph^2)^{1/2} -\arccos\aleph  \right], \nonumber\\
&&\aleph = 1-A^{-1}(\omega_H-\omega_L).
\end{eqnarray}
For a narrow hysteresis loop, $(\omega_H-\omega_L)/A\ll 1$, the offset is small, it scales as $[(\omega_H-\omega_L)/A]^{3/2}$. In the central part of the region $0 < \omega_H-\omega_L < 2A$ the offset weakly depends on $A$ and is almost linear in $\omega_H-\omega_L$. As $\omega_H - \omega_L$ approaches $2A$ the offset smoothly approaches the small-$A$ value $-AC_V\Delta S$, which applies to the leading order in $A$ for  $\omega_H-\omega_L > 2A$. The reduction in offset compared to the small-$A$ limit is easy to understand,  since going round the hysteresis loop for $2A > \omega_H-\omega_L$ the system samples both absorption branches in each modulation cycle. Both theoretically and experimentally, the overall negative offset increases with increasing microwave power, as both the height of the absorption jump $\Delta S$ and the width of the hysteresis loop increase.

\section{Conclusions}

This paper describes the theory and the experimental observation of the long-sought intrinsic optical bistability and hysteresis in a quasi two-dimensional electron system. The effect is a result of strong electron correlations and the hierarchy of the relaxation times, where the electron-electron momentum and energy exchange is the fastest process, followed by a much slower momentum relaxation due to scattering by the random short-range potential of ripplons or helium vapor atoms, followed by still much slower energy relaxation due to electron energy exchange with the environment. As a result of this hierarchy, a moderately strong resonant excitation of the transitions $|1\rangle \to |2\rangle$  between the lowest subbands of motion normal to the surface leads to redistribution of the electrons over the subbands. This causes absorption bleaching due to depletion of the lowest electron subband. It occurs for much smaller power than that required for absorption saturation. 

Because of the electron-electron interaction, the redistribution of the electrons over the subbands leads to a change of the frequency $\omega_{21}$ of the  $|1\rangle \to |2\rangle$ transition. This causes bistability of the irradiated system. For a given radiation frequency $\omega_F$ the frequency $\omega_{21}$ can be ``tuned" close to $\omega_F$ as a result of  relatively strong electron population of excited subbands. Then absorption is indeed strong and the excited subband population is self-consistently significant. Alternatively, the resonance can be not that good, the electrons absorption is not that strong, the electrons mostly stay in the lowest subband, and $\omega_{21}$ is self-consistently detuned from $\omega_F$. The developed microscopic theory of the many-electron system provides a full account of this behavior.

Experimentally, for electrons on $^3$He we observed hysteresis of microwave absorption. The observation was made by measuring the magnetoconductivity in the Corbino geometry. The data refer to the regime of strong electron correlations. Different temperatures were studied, which made it possible to investigate the scattering primarily by the helium vapor atoms and by ripplons. The data for different radiation power show the increase of the hysteresis loop with the power.  A quantitative agreement is obtained between the data and the theory of the electron-electron interaction induced change of $\omega_{21}$ as function of the electron temperature. 

We have also observed absorption hysteresis by directly measuring the transmitted microwave power near the resonant frequency $\omega_{21}$, which is modulated though the applied holding field. The hysteresis is manifested as a characteristic offset of the integrated lineshape in the in-phase component of the modulated response and by a distinctive quadrature component at finite modulation amplitude.

The results of this paper provide a new insight into the role of the electron-electron interaction in strongly correlated systems and present new types of resonant nonlinear phenomena in quasi two-dimensional electron systems.

\begin{acknowledgements}
We acknowledge valuable discussions with E. Collin, P. M. Platzman, and D. Ryvkine.  DK and KK were supported in part by KAKENHI; MID was supported in part by the NSF through grant No. EMT/QIS 0829854.
\end{acknowledgements}


\begin{thebibliography}{}

\bibitem{Cole1970} M.~W. Cole, Phys. Rev. B {\bf 2}, 4239 (1970).

\bibitem{Andrei_book} {\it Electrons on Helium and Other Cryogenic Substrates}, edited by E. Y. Andrei (Kluwer Academic, Dordrecht, 1997).

\bibitem{Monarkha_book} Y. Monarkha and K. Kono, {\it Two-Dimensional Coulomb Liquids and Solids} (Springer, Berlin, 2004).

\bibitem{Grimes1974} C.~C. Grimes and T. R. Brown, Phys. Rev. Lett. {\bf 32}, 280 (1974).

\bibitem{Collin2002} E. Collin, W. Bailey, P. Fozooni, P.~G. Frayne, P. Glasson, K. Harrabi, M. J. Lea, and G. Papageorgiou, Phys. Rev. Lett. {\bf 89}, 245301 (2002).

\bibitem{Lambert1980} D.~K. Lambert and P. L. Richards, Phys. Rev. Lett. {\bf 44}, 1427 (1980); Phys. Rev. B {\bf 23}, 3282 (1980).

\bibitem{Isshiki2007} H. Isshiki, D. Konstantinov, H. Akimoto, K. Shirahama, and K. Kono, J. Phys. Soc. Jpn. {\bf 76}, 094704 (2007).

\bibitem{Volodin1981} A.~P. Volodin and V.~S. Edel'man, Sov. Phys. JETP {\bf 54}, 198 (1981).

\bibitem{KonstantinovJLTP2007} D. Konstantinov, H. Isshiki, H. Akimoto, K. Shirahama, and K. Kono, J. Low Temp. Phys. {\bf 148}, 187 (2007).

\bibitem{Konstantinov2007} D. Konstantinov, H. Isshiki, Yu. Monarkha, H. Akimoto, K. Shirahama, and K. Kono, Phys. Rew. Lett. {\bf 98}, 235302 (2007).

\bibitem{Monarkha2007} Yu.~P. Monarkha, D. Konstantinov, and K. Kono, J. Phys. Soc. Jpn. {\bf 76}, 0124702 (2007).

\bibitem{KonstantinovJPSJ2008} D. Konstantinov, H. Isshiki, Yu. Monarkha, H. Akimoto, K. Shirahama, and K. Kono, J. Phys. Soc. Jpn. {\bf 77}, 034705 (2008).

\bibitem{KonstantinovJLTP2008} D. Konstantinov, Y. Monarkha, and K. Kono, J. Low Temp. Phys. {\bf 150}, 230 (2008).

\bibitem{KonstantinovLTP2008} D. Konstantinov, Yu. Monarkha, and K. Kono, Low Temp. Phys. {\bf 34}, 377 (2008).

\bibitem{Platzman1999} P. M. Platzman and M.~I. Dykman, Science {\bf 284}, 1967 (1999).

\bibitem{Dykman2003} M.~I. Dykman, P.~M. Platzman and P. Seddinghrad, Phys. Rev. B {\bf 67}, 155402 (2003).

\bibitem{Allen1976} S. J. Allen, D. C. Tsui and B. Vinter, Solid State Commun. {bf 20}, 425 (1976).

\bibitem{Ando1977} T. Ando, J. Physik B {\bf 26}, 263 (1977).

\bibitem{Dykman1979}M.~I. Dykman and L.~S. Khazan, Zh. Eksp. Teor. Fiz. {\bf 77}, 1488 (1979) [Sov. Phys. JETP {\bf 50}, 747 (1979)].

\bibitem{Dykman1993} M.~I. Dykman, M.~J. Lea, P. Fozooni, and J. Frost, Phys. Rev. Lett. {\bf 70}, 3975 (1993).

\bibitem{Monarkha1997}   Yu~P. Monarkha, S. Ito, K. Shirahama, and K. Kono, Phys. Rev. Lett. {\bf 78} 2445 (1997).

\bibitem{Lea1998} M.~J. Lea and M.~I. Dykman, Physica B {\bf 251}, 628 (1998).

\bibitem{Glasson2004} P. Glasson, E. Collin, P. Fozooni, P. G. Harrabi, W. Bailey, G. Papageorgiou, Y. Mukharsky, M. J. Lea, Physica E {\bf 22}, 761 (2004).

\bibitem{Bloembergen1954} N. Bloembergen and S. Wang, Phys. Rev. {\bf 93}, 72 (1954).

\bibitem{Anderson1955} P.~W. Anderson and H. Suhl, Phys. Rev. {\bf 100}, 1788 (1955).

\bibitem{Weiss1958} M. T. Weise, Phys. Rew. Lett. {\bf 1}, 239 (1958).

\bibitem{Vasilyev2002} S. Vasilyev, J. Jarvinen, A. I. Safonov, A. A. Kharitonov, I. I. Lukashevich, and S. Jaakkola, Phys. Rew. Lett. {\bf 89}, 153002 (2002).

\bibitem{Newson1987} D. J. Newson and A. Kurobe, Appl. Phys. Lett. {\bf 51}, 1670 (1987).

\bibitem{Zaluzny1993} M. Za{\l}u\.{z}ny, Phys. Rev. B {\bf 47}, 3995 (1993).

\bibitem{Craig1996} K. Craig, B. Galdrikian, J. N. Heyman, A. G. Merkelz, J. B. Williams, M. S. Sherwin, K. Campman, P. F. Hopkins, and A. C. Gossard, Phys. Rev. Lett. {\bf 76}, 2382 (1996).

\bibitem{Lutgen1996} S. Lutgen, R. A. Kaidl, M. Woerner, T. Elsaesser, A. Hase, H. Kunzel, Phys. Rev. B {\bf 54}, R17343 (1996).

\bibitem{Shtrichman2001} I. Shtrichman, C. Metzer, E. Ehrenfreund, D. Gersoni, K. D. Maranowski, and A. C. Gossard, Phys. Rev. B {\bf 65}, 035310 (2001).

\bibitem{Muller2004} T. M\"{u}ller, W. Parz, G. Strasser, and K. Unterrainer, Phys. Rev. B {\bf 70}, 155324 (2004).

\bibitem{Stockman1993} M.~I.~Stockman, L.~N.~Pandey, L.~S.~Muratov, and Th.~F.~George, Phys. Rev. B \textbf{48}, 10966 (1993)

\bibitem{Batista2002} A.~A.~Batista, P. I. Tamborenea, B. Birnir, M. S. Sherwin, and D. S. Citrin, Phys. Rev. B \textbf{66}, 195325 (2002)

\bibitem{Wijewardane2004} H.~Wijewardane and C.~A.~Ullrich, Appl. Phys. Lett. \textbf{84}, 3984 (2004);

\bibitem{Li2007} J.~H.~Li, Phys. Rev. B \textbf{75}, 155329 (2007).

\bibitem{KonstantinovPRL2009} D. Konstantinov, M. I. Dykman, M. L. Lea, Yu. Monarkha, and K. Kono, Phys. Rev. Lett. {\bf 103}, 096801 (2009).

\bibitem{Dykman1997} M.~I. Dykman, C. Fang-Yen, and M.~J. Lea, Phys. Rev. B {\bf 55}, 16249 (1997).

\bibitem{Fang1997} C. Fang-Yen, M.~I. Dykman, and M.~J. Lea, Phys. Rev. B {\bf 55}, 16272 (1997).

\bibitem{Ryvkin} D. Ryvkine, M.J. Lea, and M. I. Dykman, Abstract of the APS March Meeting (2006), http://meeting.aps.org/Meeting/MAR06/Event/45165; Invited Presentation at the Meeting on "Floating Electrons on Helium for Quantum Computing", paris (2006), http://www.princeton.edu/~lyon/paris$\_$2006.

\bibitem{Ando1978} T.~Ando, J. Phys. Soc. Jpn. {\bf 44}, 765 (1978).


\bibitem{Schuster2010}D. I. Schuster, A. Fragner, M. I. Dykman, S. A. Lyon, and R. J. Schoelkopf, Phys. Rev. Lett. {\bf 105}, 040503 (2010).

\bibitem{Shirahama1995} K. Shirahama, S. Ito, H. Suto, and K. Kono, J. Low Temp. Phys. {\bf 101}, 439 (1995).

\bibitem{PlatzmanTzoar1976} P. M. Platzman and N. Tzoar, Phys. Rev. B {\bf 13}, 3197 (1976).

\bibitem{KonstantinovJLTP2010} D. Konstantinov and K. Kono, J. Low Temp. Phys. {\bf 158}, 324 (2010).


\end{thebibliography}
\end{document}